\smallskip
\pageno=0
\font\rm=cmr12
\font\smallrm=cmr10
\font\it=cmti12

\baselineskip=20pt
\rm

\noindent \hfil Why Quantum Mechanics is Complex
\bigskip
\noindent \hfil James T. Wheeler
\smallskip
\noindent \hfil {\it {Department of Physics, Utah State University, Logan,
UT 84322}}

\noindent \hfil jwheeler@cc.usu.edu
\bigskip

\bigskip
\bigskip
\noindent \hfil Abstract

The zero-signature Killing metric of a new, real-valued, 8-dimensional
gauging of the conformal group accounts for the complex character of
quantum mechanics.  The new gauge theory gives manifolds which generalize
curved, relativistic phase space.  The difference in signature between the
usual momentum space metric and the Killing metric of the new geometry
gives rise to an imaginary proportionality constant connecting the
momentumlike variables of the two spaces.  Path integral quantization
becomes an average over dilation factors, with the integral of the Weyl
vector taking the role of the action.  Minimal $U(1)$ electromagnetic
coupling is predicted.

\vfil
\break

\overfullrule=0pt
\baselineskip=28pt

\noindent \hfil {\bf {WHY QUANTUM MECHANICS IS COMPLEX}}
\bigskip
\noindent \hfil JAMES T. WHEELER

\noindent \hfil {\it {Department of Physics, Utah State University, Logan,
UT 84322}}
\noindent \hfil jwheeler@cc.usu.edu
\bigskip

One of the more puzzling aspects of quantum mechanics is its seemingly
necessary reliance on complex quantities.  The theory's interference
effects, probability amplitudes and coupling to unitary gauge interactions
all make important use of complex numbers.  In this essay we show how a
real-valued 8-dimensional geometry can account for these behaviors in a
natural way.

This geometry, called biconformal space, arises
as a new gauging of the conformal group [1-3].  The gauging is
accomplished in three steps.  First, counting fixed points
identifies {\it {eight}}
of the conformal transformations as translations, with the remaining
homogeneous Weyl transformations (Lorentz transformations and
 dilations) forming the isotropy subgroup, ${\cal {C}}_{0}$.
Second, we construct an elementary geometry as the quotient,
 ${\cal {C}}/ {\cal {C}}_{0}$, of the conformal group,
$\cal {C}$, by ${\cal {C}}_{0}$.  This produces a principal fiber
bundle over an 8-dimensional manifold (various topologies are allowed).
Finally, we generalize to
 a curved Cartan connection by adding horizontal curvature 2-forms to the
group structure equations.  The curvature breaks both the translational
and inverse translational symmetries while retaining all 15 gauge
fields.  This contrasts sharply with previous 4-dim
gaugings of the conformal group [4 - 9], for which the inverse translational
 gauge fields are always auxiliary [5, 9].

To implement these steps we choose the $O(4,2)$ representation of the conformal
group, with connection 1-forms $\omega_{B}^{A}\enskip (A, B, \dots = 0, 1,
\ldots, 5)$.  Letting boldface or Greek symbols denote forms and $(a, b,
\ldots) = (1, \ldots, 4)$, the $O(4,2)$ metric is given by $\eta_{ab} =
diag(1, 1, 1, -1)$ and $\eta_{05} = \eta_{50} = 1$ with all other
components vanishing.  The covariant constancy of $\eta_{AB}$ reduces the
independent components of $\omega_{B}^{A}$ into four Weyl-invariant parts:
the {\it {spin connection}}, $\omega_{b}^{a}$, the {\it {solder form}},
$\omega_{0}^{a}$, the {\it {co-solder form}}, $\omega_{a}^{0}$, and the
{\it {Weyl vector}}, $\omega_{0}^{0}$ where the spin connection satisfies
$$\omega_{b}^{a} = - \eta_{bc} \eta^{ad} \omega_{d}^{c}$$
The remaining components of $\omega_{B}^{A}$ are given in terms of these.
We restrict $(A, B, \dots) = (0, 1, \ldots, 4)$ in all subsequent
equations.  The Maurer-Cartan structure equations of the conformal group
are simply
$${\bf {d\omega}}_{B}^{A} = {\bf {\omega}}_{B}^{C} \wedge {\bf
{\omega}}_{C}^{A}$$
Since no finite translation can reach the point at infinity and no inverse
translation can reach the origin, the space ${\cal {C}}/{\cal {C}}_{0}$
gives a copy of (noncompact) Minkowski space for each of the two sets of
translations.  Since the generators of the two types of translation commute
modulo the homogeneous Weyl group, the base manifold is simply the
Cartesian product of these two Minkowski spaces.  The generalization to a
curved base space is immediate.  Adding curvature 2-forms to the structure
equations and breaking them into parts based on homogeneous Weyl
transformation properties, we have:
$$\eqalignno{
{\bf {d\omega}}_{b}^{a} &= \omega_{b}^{c} \wedge \omega_{c}^{a} +
\omega_{b}^{0} \wedge \omega_{0}^{a} - \eta_{bc}\eta^{ad} \omega_{d}^{0}
\wedge \omega_{0}^{c} + \Omega_{b}^{a} \cr
{\bf {d\omega}}_{0}^{a} &= \omega_{0}^{0} \wedge \omega_{0}^{a} +
\omega_{0}^{b} \wedge \omega_{b}^{a} + \Omega_{0}^{a} \cr
{\bf {d\omega}}_{a}^{0} &= \omega_{a}^{0} \wedge \omega_{0}^{0} +
\omega_{a}^{b} \wedge \omega_{b}^{0} + \Omega_{a}^{0} \cr
{\bf {d\omega}}_{0}^{0} &= \omega_{0}^{a} \wedge \omega_{a}^{0} +
\Omega_{0}^{0}  &  (2.12) \cr} $$
The four curvatures $\Omega_{b}^{a}, \Omega_{0}^{a}, \Omega_{a}^{0}$ and
$\Omega_{0}^{0}$ are called the Riemann curvature, torsion, co-torsion and
dilational curvature, respectively.  If we set $\omega_{a}^{0} =
\omega_{0}^{0} = \Omega_{a}^{0} = \Omega_{0}^{0} = 0$ we recover a 4-dim
spacetime with Riemannian curvature $\Omega_{b}^{a}$ and torsion
$\Omega_{0}^{a}$.  If we set only $\omega_{a}^{0} =  \Omega_{a}^{0} = 0$,
we have a 4-dim Weyl geometry.

Horizontality yields curvatures of the form
$$\Omega_{B}^{A} = {\scriptstyle {1 \over 2}} \> \Omega_{Bcd}^{A} \>
\omega_{0}^{c} \wedge \omega_{0}^{d} + \Omega_{Bd}^{Ac} \> \omega_{0}^{d}
\wedge \omega_{c}^{0} + {\scriptstyle {1 \over 2}} \> \Omega_{B}^{Acd}\>
\omega_{c}^{0} \wedge \omega_{d}^{0} $$
Based on the correspondence principle relating biconformal space and phase
space ([1], [2]), $\Omega^{A}_{Bcd}$ is called the spacetime term,
$\Omega^{Ac}_{Bd}$ the cross term and $\Omega^{Acd}_{B}$ the momentum term
of $\Omega^{A}_{B}$.  The fiber symmetry group does not mix the spacetime,
cross or momentum terms.

For a {\it {flat}} biconformal space, with $\Omega_{B}^{A} = 0$, the
connection may be put into the form ([1], [2])
$$\eqalignno{
{\bf {\omega}}_{0}^{0} &= \alpha_{a}(x) {\bf {d}}x^{a} - y_{a} {\bf {d}}
x^{a} \equiv W_{a} {\bf {d}} x^{a} &(8a) \cr
{\bf {\omega}}_{0}^{a} &=  {\bf {d}} x^{a} &(8b) \cr
{\bf {\omega}}_{a}^{0} &=  {\bf {d}} y_{a} - (\alpha_{a,b} + W_{a}W_{b} -
{\scriptstyle {1 \over 2}} W^{2} \eta_{ab}) {\bf {d}} x^{b} &(8c) \cr
{\bf {\omega}}_{b}^{a} &=  ( \delta_{d}^{a} \delta_{b}^{c} - \eta^{ac}
\eta_{bd}) W_{c} {\bf {d}} x^{d} &(8d) \cr} $$
where $(x^{a}, y_{a})$ are eight independent (global) coordinates on the
space and $\alpha_{a,b} \equiv {\partial \alpha_{a} \over \partial x^{b}}$
denotes the partial of $\alpha_{a}$ with respect to $x^{b}$.

Our current discussion centers on the metric structure of biconformal
spaces.  Although biconformal space is based upon the conformal group of
Minkowski space, it does not inherit the Minkowski metric, $\eta_{ab}$.
Nonetheless, every biconformal space has a natural metric based on the
Killing metric.  The Killing metric is built from the conformal group
structure constants as
$$K_{AB} = C^{C}_{AD} C^{D}_{BC}$$
where the labels
$$(A, B, \ldots) \in \{ {\scriptstyle {{ {b\choose a}, {a\choose 0} ,
{0\choose a} , {0\choose 0}}}}  \}.$$
refer to the Lorentz, translation, inverse translation and dilatation
generators, respectively.  A short calculation gives
$$\eqalignno{
K_{AB}^{(conf.)} &= \pmatrix{
K_{\scriptstyle {{b\choose a}{d\choose c}}}^{L} & 0&0 &0 \cr
0&0 & K_{\scriptstyle {{0\choose a}{b\choose 0}}} &0  \cr
0& K_{\scriptstyle {{a\choose 0}{0\choose b}}} &0&0\cr
0&0 &0 & 1} \cr
 \cr}$$
where
$$K_{\scriptstyle {{b\choose a}{d\choose c}}}^{L} = 2(\delta^{b}_{c}
\delta_{a}^{c}  - \eta^{bd} \eta_{ac})  $$
is the Killing metric for the Lorentz group, and
$$\eqalignno{
K_{\scriptstyle {{0\choose a}{b\choose 0}}} &= \delta_{a}^{b} \cr
K_{\scriptstyle {{a\choose 0}{0\choose b}}} &= \delta_{b}^{a} \cr } $$
We now construct a metric on the bundle by expanding $\omega^{A}$ in a
coordinate basis
$$\omega^{A} = (\omega^{a}_{b}, \omega^{a}_{0}, \omega^{0}_{a},
\omega^{0}_{0}) = \omega_{\enskip M}^{A} {\bf {d}}x^{M}$$
then defining metric components
$$ g_{MN} = K_{AB} \omega_{\enskip M}^{A} \omega_{\enskip N}^{B}$$

Unlike the degenerate Killing metric for the Poincar{\' {e}} group,
$K_{AB}^{(conf.)}$ is nondegenerate both on the bundle, and when projected
to the base space.  Restricting the indices to the range $M', N' \in \{
{\mu\choose} , { \choose \nu}  \}; \enskip x^{M'} = (x^{\mu}, y_{\nu})$ the
biconformal metric is
$$\eqalignno{
g_{M'N'}  &= K_{A'B'} \omega_{\enskip M'}^{A'} \omega_{\enskip N'}^{B'} \cr
&=   \omega_{0 M'}^{a} \omega_{a N'}^{0} + \omega_{a M'}^{0} \omega_{0
N'}^{a} \cr } $$
Understanding the importance of $K_{A'B'}$ first requires understanding the
phase space correspondence principle.

In [1], [2] it is shown that dilationally flat ($\Omega_{0}^{0} = 0$)
biconformal spaces are generalizations of phase
space, with the line integral of the Weyl vector, $\omega_{0}^{0}$,
equal to the action.  The class of superhamiltonian
hypersurfaces, ${\cal {H}}(H, {\bf {p}}, x^{a}) = 0$ in the $\alpha = 0$
flat biconformal space are in 1-1 correspondence with Hamiltonian systems.
 The biconformal structure equations imply the existence of
curves satisfying Hamilton's equations.  Generalizing these dynamical
 equations by letting $\alpha \ne 0$ predicts the Lorentz force law
 for a charged particle in a background electromagnetic field,
with $\alpha = - {q \over \hbar c}A_{a}$.  This last result rescues
Weyl's geometric theory of electromagnetism [10 - 12].
Weyl equated the Weyl vector to the electromagnetic vector potential,
 but this results in predictions of unphysical size change.  By contrast,
equating $\alpha$ to the vector potential predicts the Lorentz force law
and no size change.

But the structure equations of biconformal space go beyond the usual
structure of phase space, giving a differential system for the eight solder
forms, the spin connection and the Weyl vector.  The Killing metric
described above also represents new structure, importantly different from
the usual metric on phase space.

Consider this metric difference in detail.  The usual {\it {phase space}}
metric follows from the metric on spacetime and the fact that the
4-momentum is proportional to the tangent vector to a curve.  This implies
the {\it {same}} Minkowski metric on both the configuration and momentum
parts of phase space:
$$g^{\smallrm {Ph. Sp.}}_{AB} = (\eta_{\mu \nu}, \eta^{\mu \nu})$$
The two parts of $g^{\smallrm {Ph. Sp.}}_{AB}$ are not an 8-dim metric --
they are the same metric applied to either the spacetime or the momentum
sector.

The essential point we wish to make is that the Killing metric $K_{A'B'}$
on biconformal space has eigenvalues $\pm 1$ and zero signature.
Therefore, if the inner product of the spacetime coordinates $x^{\mu}$ are
found (locally) using the Minkowski metric then the $y_{\mu}$ coordinates
{\it {must}} be contracted using {\it {minus}} the Minkowski metric
$$g^{\smallrm {Biconf.}}_{AB} = (\eta_{\mu \nu}, - \eta^{\mu \nu})$$
This is the only way to achieve zero signature while keeping the usual
Minkowski inner product on spacetime.  There is a new reason that the two
parts of the 8-dim metric must refer to two independent 4-dim metrics.
Because we have introduced a {\it {dimensionful}} metric, the scaling
weight of the spacetime and momentum parts are opposite.  The two parts
cannot be added together and must be regarded as applying only on
appropriate submanifolds.

The sign difference between $g^{\smallrm {Ph. Sp.}}_{AB}$ and $g^{\smallrm
{Biconf.}}_{AB}$ has an important consequence.  When identifying
biconformal coordinates $(x^{\mu}, y_{\nu})$ with phase space coordinates
$(x^{\mu}, p_{\nu})$, we naturally set $y_{\nu} = \beta p_{\nu}$.  The
constant $\beta$ must account for the sign difference in
$$\eta^{\mu \nu} \beta p_{\mu} \beta p_{\nu} = (-\eta^{\mu \nu}) y_{\mu}
y_{\nu}$$
$\beta$ is therefore imaginary.  Also, $\beta$ must account for the
different units of $y_{\nu}$ (length$^{-1}$) and $p_{\nu}$ (momentum), with
$\hbar$ the obvious choice.  Therefore (up to a real dimensionless
constant) we can set
$$y_{\nu} = {i \over \hbar} p_{\nu}$$
{\it {This relationship between the geometric variables of conformal gauge
theory and the physical momentum variables is the source of complex
quantities in quantum mechanics.}}

There are two further points to be discussed in identifying the proper
correspondence between the physical variables of phase space and geometric
variables of biconformal space.

First, $\eta_{ab}$ does not respect the scale invariance of biconformal
space as $K_{A'B'}$ does -- it scales with weight $+2$, and there is no
principled way to introduce it into a {\it {general}} biconformal space.
The use of diagonalizing coordinates
$$\eqalignno{
{\bf {u}}^{a} &= {\scriptstyle {1 \over \sqrt{2}}} (\omega^{a}_{0} +
\eta^{ab} \omega_{b}^{0}) \cr
{\bf {v}}_{a} &= {\scriptstyle {1 \over \sqrt{2}}} (\omega_{a}^{0} -
\eta_{ab} \omega^{b}_{0}) \cr }$$
is possible nonetheless because this transformation is {\it {symplectic}},
preserving
$${\bf {d\omega}}_{0}^{0} = {\bf {\omega}}_{0}^{a} \wedge {\bf
{\omega}}_{a}^{0}$$
The closed, nondegenerate ${\bf {d}}\omega_{0}^{0}$ corresponds to the
symplectic form in phase space.  Therefore, while it is not generally
natural to introduce a dimensionful metric into biconformal spaces, there
is no reason not to in the symplectic, flat case.

The second issue is the {\it {existence}} of spacetime as an integrable
4-dim submanifold of flat biconformal space. While the flat structure
equations {\it {are}} involute, we must ask if the involution still exists
in the ${\bf {u}}^{a}, {\bf {v}}^{a}$ basis.  Substituting into the
structure equation for $\omega^{a}_{0}$ we find
$$\eqalignno{
{\bf {du}}^{a}&={\scriptstyle {1 \over \sqrt{2}}} {\bf {d}}(\omega^{a}_{0}
+ \eta^{ab} \omega_{b}^{0}) \cr
&= {\bf {u}}^{b} \omega_{b}^{a} + \eta^{ab} {\bf {v}}_{b} \omega_{0}^{0}
\cr }$$
where we have used the antisymmetry of the spin connection,
$\eta^{ab}\omega_{b}^{c} = - \eta^{cb}\omega_{b}^{a}$.  Therefore, ${\bf
{u}}^{a}$ is in involution iff $\omega_{0}^{0} = W_{a}{\bf {u}}^{a}$ so
that the 4-dim character of $\omega_{0}^{0}$ emerges with the spacetime
manifold, which exists whenever the ${\bf {u}}^{a} = 0$ foliation is
regular.  Interestingly, this involution reduces the rewritten ${\bf
{u}}^{a} = 0$ structure equations to those of a constant curvature momentum
4-space with radius on the order of the Compton wavelength.

The standard formulation of quantum mechanics may now be seen to correspond
to a statement about measurement in biconformal space using {\it {real}}
variables $(x^{a}, y_{a})$.  In particular, the form of the Weyl vector for
a large class of solutions [1 - 3] including the flat case above is
$$\omega_{0}^{0} = \alpha_{a}(x) {\bf {d}}x^{a} - y_{a} {\bf {d}} x^{a} = -
({q \over \hbar c} A_{a}(x) {\bf {d}}x^{a} + y_{a}) {\bf {d}} x^{a}$$
Identifying the integral of $\omega_{0}^{0}$ as the action and expressing
$y_{a}$ in terms of the momentum we find
$$\int  \omega_{0}^{0} = -{i \over \hbar} \int{(p_{a} - i {q \over c}
A_{a}){\bf {d}} x^{a}}$$
We therefore correctly find both the proper minimal $U(1)$ coupling to the electromagnetic field, and the proper complex coefficient for the exponent.  The a
verage over $exp (-{i \over \hbar}{\cal {S}})$ becomes an average over $exp
(\int  \omega_{0}^{0})$.  But $exp (\int_{C}  \omega_{0}^{0})$ is precisely
the dilation factor along the path $C$, so the path integral amounts to an
averaging of size-change factors for different paths.

The reason that these two path averages can give the same results is the
difference in signature between phase space and biconformal space.  Recall
how, in the early days of relativity, the coordinate $\tau = ict$ was used
to give spacetime a Euclidean signature.  Eventually it was found easier
and more general to use an indefinite metric.  Here we see the same effect
in phase space, where our standard assumption of the same inner product on
the configuration and momentum sectors of the space has meant that we are
effectively using a purely imaginary coordinate, $p_{a} = -i \hbar y_{a}$.
By changing the signature, we can eliminate the ``i'', and work in real,
geometric variables.
\bigskip
\vfil
\break

\vfil
\break
\bigskip
\hfil References
\medskip
\item{[1]} Wheeler, J. T., Proceedings of the Seventh Marcel Grossman
Meeting on  General Relativity , R. T. Jantzen and G. M. Keiser, editors,
World Scientific, London (1996) pp 457-459.
\item{[2]} Wheeler, J. T., New conformal gauging and the electromagnetic
theory of Weyl (1996), submitted for publication,
http://xxx.lanl.gov/abs/hep-th/9706214
\item{[3]} Wheeler, J. T., Normal biconformal spaces,
http://xxx.lanl.gov/abs/hep-th/9706215, submitted for publication.
\item{[4]} Ferber, A. and P.G.O. Freund, Nucl. Phys. {\bf {B122}} (1977) 170.
\item{[5]} Kaku, M., P.K. Townsend and P. Van Nieuwenhuizen, Phys. Lett.
{\bf {69B}} (1977) 304.
\item{[6]} Crispim-Rom{\~a}o, J., A. Ferber and P.G.O. Freund, Nucl. Phys.
{\bf {B126}} (1977) 429.
\item{[7]} F. Mansouri, Phys. Rev. Lett. 42 (1979) 1021.
\item{[8]} F. Mansouri and C. Schaer, Phys. Lett. 101B (1981) 51.
\item{[9]} Wheeler, J. T., Phys Rev D{\bf {44}} (1991) 1769.
\item{[10]} Weyl, H., Sitzung. d. Preuss. Akad. d. Wissensch. (1918) 465.
Reprinted in:  The Principle of Relativity, Chapter XI, (Dover, 1923) 199
-216.
\item{[11]} Weyl, H., {\it {Space-Time-Matter}}, Dover Publications, New
York (1952).  Originally:  H. Weyl,  Raum-Zeit-Materie, (3rd Ed.,1920),
Chapts II \& IV, §§34+35, p.242,et seq.
\item{[12]} Additional works by Weyl on the subject of geometry and
electromagnetism include Weyl, H., Relativity Symposium, Nature 106
781(1921); Math. Zeitschr.  2 (1918) 384; Ann. d. Physik,  54  pp117
(1918); Ann. d. Physik,  59 (1919) 101; Phys. Zeitschr. 21 (1920) 649; Ann.
d. Phys. 65 541(1921); Phys. Zeitschr. 22 (1921) 473; Zeit. f. Physik, 56
(1929) 330.

\bye